# MULTI-FREQUENCY MONITORING OF THE SEYFERT 1 GALAXY NGC 4593 II: A SMALL, COMPACT NUCLEUS ?

M. SANTOS-LLEÓ[1 2], J. CLAVEL[3], P. BARR[3],I.S. GLASS[4], D. PELAT[1], B.M.PETERSON[5],G. REICHERT[6]

[1]*Observatoire de Paris, 92195 Meudon Principal Cedex, France*
[2]*ESA IUE Observatory, P.O. Box 50727, 28080 – Madrid, Spain*
[3]*ISO Observatory, Code SAI, ESTEC, Postbus 299, 2200 AG Noordwijk, The Netherlands*
[4]*South African Astronomical Observatory, P.O. Box 9, Observatory, 7935 South Africa*
[5]*Department of Astronomy, Ohio State University, 174 West 18th Avenue, Columbus, OH 43210, USA*
[6]*NASA-Goddard Space Flight Center, Code 684.9, Building 21, Greenbelt, MD 20771*

## ABSTRACT

We discuss the results of a campaign to monitor spectral variations in the low-luminosity Seyfert 1 galaxy NGC 4593, at X-rays, ultraviolet, optical and near IR frequencies. The observations and data analysis have been described in a companion paper (Santos-Lleó et al. 1994; Paper 1). The active nucleus in this galaxy is strongly and rapidly variable in all wavebands, implying that the continuum source is unusually compact. Its energy distribution from 1.2 $\mu$m to 1200 Å obeys a power-law whose index is significantly steeper than is usual in Seyfert's or QSO's; the "big bump" is either absent or shifted to wavelengths shorter than 1200 Å. The variations of the soft-X ray *excess* do not correlate with those of the UV or hard X-ray continuum. The far UV and optical fluxes are well correlated, while the correlation between the hard X-rays and 1447 Å continuum is only marginally significant. Moreover, the optical flux cannot lag behind the UV by more than 6 days. These results cannot be accommodated in the framework of the standard geometrically thin accretion disk model. Rather, they suggest that the bulk of the UV and optical flux originates from thermal reprocessing of X-rays irradiating the disk. The soft X-ray excess is probably the only spectral component which originates from viscous dissipation inside the disk and the near infrared is probably emitted by hot dust heated by the UV radiation. Such a model is consistent with NGC 4593 having a relatively small black-hole mass of the order of $2 \times 10^6 M_\odot$ as inferred from the line variability study.

The high ionization/excitation emission lines are very broad and strongly variable and their variations correlate with those of the continuum. The low excitation lines are significantly narrower and remain constant within the accuracy of our measurements. These results suggest a stratified BLR, where the degree of ionization and the velocity dispersion of the gas increase toward small radii. The Ly$\alpha\lambda$1216 line responds to the variations of the continuum with a delay $\leq$ 4 days. To a first order approximation, the BLR in NGC 4593 is well modelled with two different zones at distances of $\sim$ 15 and 3 lt-ds from the ionizing source respectively.

## 1. INTRODUCTION

Despite a flurry of publications, active galactic nuclei are still far from being theoretically understood. It is generally believed that the release of gravitational energy by matter falling onto a massive black hole (BH) is at the origin of their tremendous luminosity. Circumstantial support for this idea comes from the fact that BH accretion disk (AD) models reproduce reasonably well the salient "big bump" which characterizes the AGN Spectral Energy Distribution (SED) in the UV-optical regime (Lynden-Bell 1969; Shields 1978; Malkan 1983). The bump is characterized by an excess of energy over the far UV to the optical–near IR wavelength range (*e.g.* Sanders et al. 1989). Typically, its spectral index, $\alpha \simeq -0.6$ ($F_\nu \propto \nu^\alpha$; O'Brien Gondhalekar & Wilson 1988; Richstone & Schmidt 1980; Cheney & Rowan-Robinson 1981), is significantly harder than that of the IR continuum and the proposed underlying IR to X-ray power-law ($\alpha \leq -1$). Extrapolation of the UV spectrum to $keV$ energies overestimates the observed X-ray flux by at least an order of magnitude. Hence, the bump must turn over somewhere in the largely unexplored EUV range. Detailed fits to the SED of large numbers of AGN's have been performed (*e.g.* Edelson & Malkan 1986). From such fits, one is in principle able to infer the most fundamental parameters of the AD theory, namely the mass of the black hole M and the accretion rate $\dot{M}$. Note that geometrically thin AD models only aim at explaining the big-bump. By their construction, they cannot account for the X-ray and $\gamma$-ray spectra, the disk being far too cool to emit significantly beyond a few hundred $eV$. Recently however, a hybrid model has been put forth by Wandel & Liang (1991) where both a cool ($T \leq 10^6 K$) optically thick and a hot ($T \sim 10^9 K$) geometrically thick optically thin phase coexist in the disk. In this unified theory, both the optical-UV and X-ray regimes are accounted for at once.

The variability of the nuclear source has the potential to yield clues to the origin of the SED. Repeated simultaneous observations in different wavebands are necessary to distinguish between competing models. Variability is useful to search for possible causal connections between different wavebands. As an example, the recent discovery of correlated variability of the optical, UV and X-ray flux in NGC 5548, with no detectable lag larger than a few days (Clavel et al. 1991 & Clavel et al. 1992). Such observations sharply constrain the range of allowed models (Malkan 1991; Courvoisier & Clavel 1991; Collin-Souffrin 1991; Krolik et al. 1991). In particular, they essentially rule-out the standard geometrically thin AD model since they imply a sound speed which is a substantial fraction of the velocity of light ($\sim 0.1c$).

The rich emission line spectrum characteristics of AGN's has also proven to be a powerful diagnostic tool. Monitoring programmes have shown that the broad permitted lines vary together with the continuum, thereby indicating that they originate in a gas which is heated and photoionized by the continuum source. Because the gas local emissivity response is quasi-instantaneous, any delay between the variations of the continuum and those of the lines is due to the finite light propagation time and therefore contains information about the average distance of the gas from the ionizing source. Also, line profiles and their variations hold clues to the kinematics of the BLR. Peterson (1993) reviews the application of this so-called "reverberation mapping" technique for determining the BLR geometry and kinematics.

In a companion paper (Santos-Lleó et al. 1994; paper 1), we presented the data obtained during a multiwavelength monitoring campaign of NGC 4593, from the near IR to the X-rays. The nuclear continuum flux was measured in selected wavelength windows together with that of the emission lines. The "pure" AGN spectrum was isolated from other parasitic contributions such as light from the dust and stars in the host galaxy and the pseudo-continuum created by Fe II lines and Balmer continuous emission. In this paper, we discus the properties of the intrinsic spectrum of the active nucleus in NGC 4593. In section 2, we investigate the properties of the nuclear energy distribution over a broad frequency range and in section 3, the pattern of variability of the SED and the relationship between the emission processes in the different wavebands. In section 4, we study the variations of the broad emission lines in relation to those of the continuum. Section 5 discusses the implication of our findings in the context of existing models. The main results are summarized in section 6.

## 2. THE IR TO X-RAYS SPECTRAL ENERGY DISTRIBUTION

In this section we investigate the instantaneous spectral energy distribution (SED) of the nuclear source in NGC 4593, from 100 $\mu$m to 10 $keV$.

### 2.1 The UV-Optical nuclear continuum

After removal of the stellar emission component (paper 1), the SED of NGC 4593 from $\sim$ 0.12 $\mu$m to 1 $\mu$m follows a power-law $F_\nu \propto \nu^\alpha$. A linear fit to the only truly contemporaneous optical-ultraviolet spectrum of February 15, 1985 yields $\alpha = -0.83\pm0.06$ (see also fig 5 in paper 1)

The amount of Fe II + BaC contamination below $\sim$ 1900 Å is quite small (paper 1, fig. 5). Hence, a flux–flux diagram at two different far-UV wavelengths also yields information on the spectral shape of the variable continuum and its temporal behaviour. Figure 1 shows the strong linear correlation ($r$=0.98 for 18 degrees of freedom (d.o.f.); $P(r \geq 0.98)=8\times10^{-12}$) that exists between $F_{1447}$ and $F_{1803}$, the continuum intensities at 1447Å and 1803Å respectively. A linear fit ($\chi^2_\nu$ = 0.71) yields:

$$F_{1803} = (0.84\pm0.05)\times F_{1447} + (0.12\pm0.10)\ (10^{-14}\mathrm{erg\ s^{-1}cm^{-2}Å^{-1}})$$

The intercept corresponds closely to the mean amount of Fe II plus Balmer continuum line emission at 1803 Å predicted by our model decomposition of the "small bump": $0.13 \pm 0.02\ 10^{-14}$erg s$^{-1}$cm$^{-2}$Å$^{-1}$ (paper 1). After removal of the Fe II+BaC emission, the intercept of the best-fit line is therefore null to within the uncertainties. This means that the intrinsic flux of the nuclear continuum at 1803 Å varies in strict proportion to $F_{1447}$ *i.e.* that the far-UV continuum retains a constant spectral shape while its intensity changes. The spectral index is readily obtained from the slope of the above best-fit regression line. After correction for reddening, one obtains: $\alpha = -1.13 \pm 0.28$. A linear relation analysis between the 1803 Å band and the 1333 Å band yields an almost identical result: $\alpha = -1.15 \pm 0.31$. Averaging the 2 results, one obtains the far-UV power-law spectral index:

$$\alpha_{fuv} = -1.14 \pm 0.21$$

The continuum spectral shape of NGC 4593 is unusually steep compared to that of most Seyfert 1 galaxies and quasars. For instance, O'Brien et al. (1988), find a mean far ultraviolet spectral index $\langle \alpha_{uv} \rangle = -0.67 \pm 0.05$ for a sample of 68 non-Blazars quasars. There are no statistical studies of Seyfert 1 galaxies, but detailed studies of individual objects tend to yield similarly flat far ultraviolet spectral index, *e.g.* $\alpha = -0.02 \pm 0.12$ in NGC 4151 (Clavel et al. 1987), $\alpha = -0.46 \pm 0.11$ in Fairall 9 (Clavel et al. 1989), and $\alpha \leq -0.7$ (average) in NGC 5548 (Clavel et al. 1991).

### 2.2 The 10 keV–100 $\mu$m energy distribution

The overall February 1985 SED of NGC 4593 from 10 *keV* to 100 $\mu$m is shown in figure 2. It has been corrected for interstellar reddening, cool interstellar dust emission in the FIR and stellar light. The Balmer continuum and Fe II line emission in the optical and near UV, are responsible for the "small bump" centered at $\nu \approx 10^{15}$ Hz. Only the ultraviolet, optical and IR observations are truly simultaneous. The 2–10 *keV* X-ray power-law represents an average over our 7 ME EXOSAT observations. The dotted line accounts for the soft X-ray flux as the sum of the extrapolation of the hard X-rays power law plus the soft excess (paper 1). With the possible exception of the far IR data, which is probably an upper limit to the true nuclear flux, figure 2 should be a fairly accurate representation of the nuclear energy distribution in NGC 4593. For convenience, we have also plotted the 20 cm flux as measured by Ulvestad & Wilson (1984).

The main characteristics of the NGC 4593's SED is its steepness and the weakness of its "big blue bump". As a matter of fact, the X-ray point lies less than a factor of 2 below an extrapolation of the optical–UV power–law, in sharp contrast with what is observed in the majority of Seyfert 1's or quasars, where the X-ray spectrum lies one to two orders of magnitude below the extrapolated ultraviolet continuum. McDowell et al. (1989) identified a new class of so-called "weak-bump" quasars, which are characterized by small (*i.e.* $\leq 0$) values of their far-ultraviolet-to-near-IR color index $C_{UV/IR} = \log L_{FUV}/L_{NIR}$, where $L_{FUV}$ is the integrated 1000–2000 Å UV luminosity and $L_{NIR}$ is the 1 to 2 $\mu$m near IR luminosity. After correction for reddening and removal of near-IR stellar light, NGC 4593's mean index averaged over all epochs of observation is $+0.0\pm0.1$, and lies in the same range as the 2 weakest–bump QSO's in McDowell's sample: $C_{UV/IR} = -0.20$ and $-0.01$, respectively. Note that the index of NGC 4593 is variable and sometimes reaches values as low as *e.g.* –0.16 at the time of our simultaneous IUE and JHKL observations of February 15, 1985.

There are *a priori* two reasons that could explain the small value of the ultraviolet-to-near-IR color index in NGC 4593: either the UV flux is depressed or the near IR flux is enhanced. To check the latter possibility, we have computed a second parameter, $\alpha_{IR-X}$, the 2-point 6 *keV* to 3.5 $\mu$m spectral index (Glass 1981; Carleton et al. 1987). The value of $\alpha_{IR-X}$ in NGC 4593 was obtained from the average of the 7 hard X-ray measurements and the average of the 20 L band measurements. This yields $\alpha_{IR-X} = -1.08 \pm 0.05$, where the error results from the propagation of the 1-$\sigma$ dispersion of the flux in each waveband due to variability. The value of $\alpha_{IR-X}$ in NGC 4593 is not significantly different from the average index obtained by Carleton et al. (1987) for their sample of "class A" *i.e.* unreddened Seyfert 1 galaxies, $\alpha_{IR-X} = -1.05 \pm 0.05$. Another way of quantifying the weakness of the bump in NGC 4593 is to compute its 1450 Å-to-2-*keV* spectral index. We obtain $\alpha_{UVX} = -0.99 \pm 0.06$ and $-1.08\pm0.03$ for the simultaneous X-ray and UV observations of February 3, 1984 (bright state) and January 10, 1986 (faint state), respectively. These indices are large compared to the mean value of –1.4 found in quasars (O'Brien et al. 1988). The optical-to-X-ray spectral index is also unusually large in NGC 4593. For instance, $\alpha_{ox} = -1.09 \pm 0.06$ in February 1984, far above the mean spectral index of radio-quiet quasars $\langle\alpha_{ox}\rangle = -1.46^{+0.05}_{-0.07}$ (Zamorani et al. 1981) or that of the 81 Seyfert 1 galaxies and quasars studied by Mushotzky and Wandel (1989): $\langle\alpha_{ox}\rangle = -1.36$.

## 3. THE PATTERN OF VARIABILITY OF THE CONTINUUM

### 3.1 Amplitudes & time scales

The light-curves of the continuum at 1447, 2710 and 5200 Å (FES) are displayed in figure 3 while those of the near-infrared J, H, K and L flux are shown in figure 4. The main characteristics of the continuum variability are summarized in table 1 where we list for each of the wavelength bands, the unweighted mean flux (in mJy) averaged over all epochs, the fractional variation $F_{var}$ (defined as the ratio of the r.m.s. fluctuations to the mean flux, after subtraction of the measurement error in quadrature), the ratio of maximum to minimum flux, $R_{max}$, the reduced chi-square $\chi^2_\nu$ (calculated under the assumption that the flux did not vary and remained equal to its average value) together with the number of degrees of freedom and the two-folding time, $\Delta t_{\times 2}$, (Penston et al. 1981) *i.e.* the time it took for the flux to increase or decrease by a factor 2. When the amplitude of the observed variation only reached a factor, f, $1.5 \leq f < 2$, the corresponding entry has been put in parenthesis to indicate that it has a lower significance.

### *3.1.1. X-Rays*

The most rapid fluctuations take place in the X-rays where the two-folding time is of the order of 1.1 hours (Barr et al. 1986). Superimposed on these rapid fluctuations are longer term ($\sim$ weeks to years), larger amplitude ($R_{max} \sim 4 - 6$), variations. The values of $F_{var}$ and $R_{max}$ in table 1 apply to the ME and LE count-rates. Adding pre-EXOSAT measurements increases only slightly $F_{var}$ to 0.46. Systematic errors in determining the conversion factors between various instruments undoubtedly introduces artificial variability in the power-spectrum so that the difference in $F_{var}$ is not significant. Hence, extending the base-line from 2 to $\sim$ 15 years, does not reveal larger amplitude variations. As a matter of fact, the full range of variability was already covered in the one month interval which separates the two 1978 HEAO1 measurements. NGC 4593 is therefore qualitatively similar to NGC 5506 whose much better sampled power-spectrum clearly flattens at frequencies $\leq (20 days)^{-1}$ (McHardy 1988). The amplitude of the variations is significantly larger at lower energies ($F_{var} \sim 0.6$ in the LE) than in the 2–10 $keV$ band ($F_{var} \sim 0.4$), as seems to be a common property of Seyfert galaxies (Grandi et al. 1992). Again, this argues for the presence of an additional and highly variable separate soft component in the X-ray spectrum of NGC 4593. After subtraction of the extrapolated ME component, the fractional variation of the soft X-ray excess is 1.00, higher than at any other frequency.

### *3.1.2. Ultraviolet*

The far-ultraviolet ($\lambda < 2000$ Å) continuum varied by a factor $R_{fuv} = 4.1 \pm 0.7$, where $R_{fuv}$ is the mean of $R_{1333}$ and $R_{1447}$. Thus, the amplitude of variability in NGC 4593 is comparable to that of NGC 5548 during the 1988–1989 campaign ($R_{fuv} = 4.0 \pm 0.4$; Clavel et al. 1991) but much smaller than that of NGC 4151 ($R_{fuv} = 11.8 \pm 1.6$; Clavel et al. 1987) or F9 ($R_{fuv} = 33 \pm 5$; Clavel et al. 1989)

The most rapid variations took place in February 1984, (JD 5733-5736) when the 1447 Å and the 1333 Å fluxes decreased by a factor 2.24 $\pm$ 0.46 and 1.94 $\pm$ 0.34 respectively in 69 hours. Averaging the results at 1447 Å and 1333 Å, yields $\Delta t_{\times 2}$ = 2.7 $\pm$ 0.8 days which sets an upper limit of $(2.8 \pm 0.8) \times 10^{16}$ cm (0.009$\pm$0.003 pc) to the "diameter" of the far-ultraviolet source (Terrell 1967). Similar two-folding times and dimensions are found in NGC 4151 (Perola et al. 1982) which has about the same absolute luminosity as NGC 4593.

A search for even faster *i.e. intra-day* variability was twice unsuccessful. For 2 epochs, 5 consecutive short wavelength spectra had been obtained over a time span of $\sim$ 16 hours (paper 1). We measured in each individual spectrum the intensity of the continuum in the 4 SW windows as well as the integral of the flux over the entire SWP spectral range (from 1226 to 1945 Å). We then compared the dispersion of those measurements with the accuracy expected from the IUE instrument. The quoted r.m.s. repeatability of the IUE flux in 25 Å bins for optimally exposed SWP spectra is 2.5 % (Bohlin, Holm & Lindler 1981). However, our spectra are not optimally exposed in the continuum. At 1803 Å, where the S/N is the highest, the ratio of the r.m.s fluctuations of the individual measurements to the mean is 6.1 % in June 1987 and 3.2 % in December 1987. After quadratically removing the measurement errors (defined now as being the error on the mean flux in the 40 Å wide 1803 Å window), these figures become 4.5% & 2.2% respectively. For the integrated SWP flux, the ratio of the r.m.s. fluctuations to the mean is 2.8 % and 2.0 % in June and December 1987 respectively. Finally, a $\chi^2_\nu$ test - where the error assigned to each individual flux is now the error on the mean in the interval of integration - also rules out, at better than the 99.9% confidence level, the existence of significant UV variations on time scales of several hours. For instance, applying the test to $F_{1447}$ yields $\chi^2_\nu = 0.14$ for 1987, June 25 and 1.1 for 1987 December 14.

The continuum variations are well correlated in all ultraviolet bands: *e.g.* the linear correlation coefficient between $F_{1447}$ and $F_{2710}$ is $r = 0.90$. For wavelengths shorter than 2000 Å, the amplitude of the variations is essentially independent of the frequency, in agreement with our earlier finding that the intercept of the best fit regression line between two far-ultraviolet bands is zero. Above 2000 Å, by contrast, the amplitude of the continuum variations decreases sharply and systematically with increasing wavelength (table 1). This implies that the spectrum appears systematically bluer when the source gets brighter or, equivalently, that the slope of the continuum is positively correlated with the flux. A single power-law does not provide a good match to the entire 1200–3200 Å range. Nevertheless, we have computed the index of such a power-law for illustrative purpose. Its mean value averaged over all epochs of observation $< \alpha_{IUE} > = -2.0$. This is significantly steeper than the spectral slope of the far UV continuum. Moreover, $\alpha_{IUE}$ varies from epoch to epoch and its variations, as expected, are strongly correlated with those of the 1447 Å flux (r=0.869, 99.999% confidence level). This effect was noted previously by Edelson, Krolik and Pike (1990) in a sample of Seyfert galaxies. They explained it as an intrinsic property of the nuclear continuum. In the case of NGC 4593 at least, a more natural explanation is dilution of the long wavelength ($\geq$ 2000 Å) pseudo-continuum flux by a quasi-constant source of emission from blended Fe II lines and Balmer continuum emission. Indeed, it is shown latter that neither the ultraviolet Fe II lines nor the Balmer continuum show appreciable variations. Such a conclusion is strengthened by the fact that the Mg II$\lambda$2798 line flux also remains constant within the measurement errors (10 %; see below). In favour of the dilution hypothesis, we note that the amount of Fe II + BaC flux at 2710 Å necessary to model the small bump of NGC 4593, 0.78±0.08 $10^{-14}$erg $s^{-1}cm^{-2}Å^{-1}$(see paper 1), corresponds closely to the intercept of the best fit (r=0.896) regression line between the 1447 & 2710 Å bands: 0.69±0.04 $10^{-14}$erg $s^{-1}cm^{-2}Å^{-1}$.

### *3.1.3. Optical*

As can be judged from table 1, the optical (FES) continuum also shows highly significant variations. However, their amplitude ($F_{var}$ = 0.08) is much smaller than in the ultraviolet. This is because the optical nuclear flux is diluted by the strong background stellar light from the underlying galaxy. After removal of the galaxy contribution (see paper 1), the fractional variation of the FES is $F_{var}$ = 0.3, slightly smaller than in the far-UV ($F_{var}$ = 0.4) This may indicate the existence of subtle variations of the optical-UV spectral index similar to those found in NGC 5548 (Clavel et al. 1992). However, our measurement errors in the optical are too large to allow us to draw a firm conclusion on that issue.

Because the measurement errors are of the same order as the amplitude of the rapid variability, we are not able to reach any definitive conclusion concerning the existence of optical variations on time scales shorter 150 day, the interval between December 31, 1979 and May 29, 1980 during which the observed FES count rate dropped by 20 % implying that the galaxy subtracted flux had decreased by a factor 2.4±0.6.

### *3.1.4. Near IR*

In the near IR, the amplitude of variations of the flux measured through a 12″ aperture is quite small. Nevertheless, even in the J band where they are the smallest, the variations are highly significant, the usual chi-square test yielding $\chi^2_\nu = 2.38$ with a probability that such a large value arises by chance $\leq 2 \times 10^{-4}$. The fractional variation of the 1.2 $\mu$m flux, is smaller than at any other wavelengths. This is because the relative contribution of the stellar population is maximum in the J band. A completely different picture emerges after removal of the stellar light (table 1). The fractional variation of the "net" – *i.e.* galaxy subtracted – flux is maximum at 1.65 $\mu$m and declines smoothly toward longer wavelength.

The most rapid statistically significant (at the 3.2 $\sigma$ level) IR variation is the 13 % decrease of the 2.2 $\mu$m flux which occurred in the 11 days between the observations of February 15 and 26, 1985 (Fig. 4). After galaxy subtraction, the relative variation becomes 30 %. The implied two-folding time is 37±12 days only, indicating that the source of the variable IR flux in NGC 4593 is unusually compact.

The fluxes in the 4 near IR bands are well correlated. However, the correlation is significantly better between J and H ($r$ = 0.88; P(r≥0.88) = 8 $10^{-8}$) or H and K ($r$ = 0.95; P(r≥0.95) = 3 $10^{-11}$) than between J and L ($r$ = 0.76; P(r≥0.76) = 1 $10^{-4}$).

### 3.2 Relationship between the various wavebands

In this section, we investigate the possible correlations, or absence thereof, between the continuum intensity in the different frequency bands. We only discuss those pairs of bands for which there exists a sufficient number of contemporaneous measurements. The results are summarized in table 2. It lists for each pair of wavebands, the correlation coefficient between the fluxes, r, the probability of exceeding r by chance and, when relevant, the reduced $\chi^2$ resulting from a linear fit to the data.

#### *3.2.1. X-Rays versus UV*

There are 4 sets of truly contemporaneous X-ray and ultraviolet observations. A fifth EXOSAT observation, obtained on February 3, 1984, is bracketed by two IUE observations made on February 2 and 5. Since on that occasion the ultraviolet continuum underwent large variations, we linearly interpolated the UV light-curve to obtain the value of $F_{1447}$ at the exact time of the X-ray pointing. Given the absence of significant high-frequency ($\geq (16\ hours)^{-1}$) power in the ultraviolet power density spectrum (PDS) of NGC 4593 (see section 3.1.2), NGC 5548 (Clavel et al. 1991) and NGC 4151 (Clavel et al. 1990), the interpolation of the UV light-curve over a $\sim$ 1.5 day time scale should provide a reasonably accurate estimate of the true ultraviolet flux at the time of the EXOSAT pointing.

As can be seen from fig 5 and table 2, both the ME and the LE count rates correlate with $F_{1447}$. Note, however, that a Spearman rank correlation test yields a 28 % probability that the correlation between the ME flux and $F_{1447}$ arises by chance, indicating that the result is only marginally significant. This result needs to be confirmed with more simultaneous observations in the two energy bands. Highly significant correlations between the X-ray and UV flux have been found in NGC 5548 (Clavel et al. 1992) and NGC 4151 (Perola et al. 1986).
Assuming the correlation is real, the best-fit ($\chi^2_\nu = 0.82$; P[$\chi^2_\nu \geq 0.82$] = 0.48) linear relation analysis between the ME flux at 4 $keV$ and $F_{1447}$ – with both quantities now expressed in $\mu$Jy – yields:

$$F_{4keV} = (2.04 \pm 0.50)\ 10^{-3} \times \mathrm{F}_{1447} + (0.11 \pm 0.33)(\mu\mathrm{Jy})$$

A very similar relation obtains for the LE flux rate versus $F_{1447}$. The intercept of the best fit line is zero within the uncertainty. This implies that the hard X-Ray flux in NGC 4593 is strictly *proportional* to the intensity of the Ultraviolet continuum. The best-fit slope can therefore be converted into a 1447 Å–4 $keV$ UVX spectral index. After correction of the UV flux for reddening, one obtains:

$$\alpha_{uvx} = -1.04 \pm 0.05$$

in good agreement with the average UV-to-X-ray spectral index inferred earlier (section 2.2). As noted earlier, $\alpha_{uvx}$ in NGC 4593 is significantly flatter than the mean UV/X-Ray spectral index of quasars (–1.4; O'Brien et al. 1988).

### *3.2.2. Optical versus UV*

As can be judged from fig. 6 and table 2, the optical FES count rate correlates well with the ultraviolet continuum flux at 1447 Å. A linear fit yields:

$$CNTS_{FES} = (14.1 \pm 1.6) \times \mathrm{F}_{1447} + (66 \pm 2)$$

where $\mathrm{F}_{1447}$ is expressed in units of $10^{-14}$erg s$^{-1}$cm$^{-2}$Å$^{-1}$. Within the uncertainty, the best-fit intercept is equal to the count rate expected from the amount of stellar light which falls into the FES aperture, 60±6 counts (see paper 1). This implies that the "net" – *i.e.* galaxy subtracted – optical flux varies in strict proportion to the ultraviolet flux. In other words, our data are consistent with the hypothesis that the overall optical-UV continuum, (with the exception of the ~ 2000–3000 Å range where it is diluted by Balmer continuum and Fe II emission; see section 3.1.2) retains a constant spectral shape while its intensity varies. Converting the slope of the best-fit regression line into a spectral index yields (after correction for reddening) $\alpha_{fes/uv} = -0.87 \pm 0.09$.

So far, we have obtained two estimates of the UV-optical spectral index (section 2.1):

- $\alpha_{15feb} = -0.83 \pm 0.06$ from the 15 February 1985 combined UV-optical spectrum
- $\alpha_{fes/uv} = -0.87 \pm 0.09$

These two estimates agree well within the uncertainty. Taking the weighted mean yields:

$$\alpha_{uvo} = -0.85 \pm 0.04$$

for the average spectral index of NGC 4593 over the optical-UV range. Moreover, the formal error on $\alpha_{fes/uv}$ allows us to set a 3 $\sigma$ upper limit on the allowed range of spectral variations and rule-out, at the 99.7 % confidence level, variations larger than:

$$\Delta\alpha_{uvo} = \pm\, 0.27$$

The spectral index found earlier for the far-UV continuum alone ($\alpha_{fuv} = -1.14 \pm 0.21$) is not significantly steeper than that of the optical-ultraviolet continuum, since the difference amounts to 1.4 $\sigma$ only. The spectral index inferred from the ultraviolet/X-ray correlation ($\alpha_{uvx} = -1.04 \pm 0.05$) is steeper than $\alpha_{uvo}$ but only at the $2.8\sigma$ level of confidence. This may indicate the existence of a gentle downward curvature of the SED, the spectrum bending gradually toward higher frequencies. Such a steepening may allow the extension of the optical-UV continuum to "connect smoothly" with the hard X-ray spectrum whose average index is $\alpha_{2-10KeV} = -0.75 \pm 0.05$.

To summarize, all available data are consistent with a fairly simple SED: a single steep power-law extending from the optical to the X-ray band with a slight downward curvature. Moreover, this SED retains a constant shape while its flux varies, at least to a first order approximation ($\Delta\alpha \leq \pm\, 0.27$). It maybe worth recalling however, that superimposed on this simple spectrum, there exists a separate component in the soft X-ray band whose variations are larger than and uncorrelated with those at other frequencies.

The only subset of the light-curves which would theoretically warrant a cross-correlation analysis is that of February 1985. However, given the large errors and the contamination of the FES nuclear counts by the underlying galaxy light, no significant FES variations could be detected at that epoch. Nevertheless, the strong correlation between the optical and ultraviolet flux together with the very short time scale of the UV variations still allows a stringent upper-limit on any delay between the variations in the two wavebands to be set. This is because the FES/UV correlation would not survive if the 5200 Å flux lagged by more than 70 hours, the two-folding time of $F_{1447}$. To test this hypothesis, we have performed a Monte-Carlo simulation: to mimic a delay of $\Delta t_{\times 2}$ between the ultraviolet and optical, artificial UV time series were generated by randomly selecting with an equal probability any flux value in the interval $0.5\times f_{obs,i}$–$2.0\times f_{obs,i}$, where $f_{obs,i}$ is the observed UV flux at epoch $i$. The experiment was repeated 100 times, and for each iteration the correlation coefficient $r$ between the FES and the artificial ultraviolet time series was computed. For only 43 % of the time does $r$ exceed the threshold of 0.56 (99 % confidence level for 18 *d.o.f.*). In other words, if the UV light-curve had been sampled 3 days earlier or 3 days later than the FES flux, there would only have been a 43 % chance of finding a significant correlation between the two bands. The above computation relies on the assumption that episodes of rapid variability are not a rare occurrence in NGC 4593. As a matter of fact, out of the 9 pairs of epochs separated by 5 days or less, $F_{1447}$ varied by more than 40 % on three occasions (*i.e.* one third of the time). We tentatively assign a conservative upper-limit of 6 days to any time delay between the ultraviolet and optical continuum. Even more stringent upper limits have been obtained for NGC 5548 ($\Delta t \leq 2$ days; Clavel et al. 1991), NGC 4151 ($\Delta t \leq 2$ days; Clavel et al. 1990) and NGC 3783 ($\Delta t \leq 1$ day; Stirpe et al. 1994).

### *3.2.3. Near IR versus UV*

The IR observations of February 1985 are close in time but not strictly simultaneous with those in the ultraviolet and the sampling is too coarse to permit a meaningful cross-correlation analysis of the two time series. However, it is obvious from a comparison of fig. 3a with fig. 4b that while the K flux decreased monotonically by about a factor 1.3, the UV continuum behaved erratically with changes of up to a factor of two. This suggests that the variations at 2.2 $\mu$m are decoupled from those in the ultraviolet (and optical), at least for any lag shorter than 8 days.

## 4. EMISSION LINES

### 4.1 Amplitude & time scales

The light-curves of the strongest emission lines are shown in figure 7. Table 3 lists the variability parameters of each line as defined in section 3.1.
Only five lines exhibit significant (at the 99.9 % level) variability: Ly$\alpha\lambda$1216, C IV$\lambda$1549, C III]$\lambda$1909, Si IV$\lambda$1394 and He I$\lambda$5876. The He II$\lambda$1640 and N IV]$\lambda$1486 variations are only significant at the 89 % and 64 % confidence level respectively. This is because these lines are weak and the relative uncertainty on their flux is therefore large. More remarkable is the absence of significant variations of the strong and well exposed Mg II$\lambda$2798 line ($\chi^2_\nu \sim$ 0.4), at least to the $\sim$ 10 % level of accuracy of our measurements. No significant variations of the Balmer lines have been detected either, probably due to the small number of observations available. Note however that low amplitude variations of the Balmer lines have been reported by Dietrich et al. (1993). The UV and optical Fe II lines did not vary either, nor did the Balmer continuum (Table 3). In short, only the high excitation/high ionization lines seem to vary in NGC 4593.

The amplitude of the variability decreases along the sequence Ly$\alpha$:CIV:CIII], the corresponding fractional variations being 0.28, 0.23 and 0.20. While these figures indicate strong line variability – with changes of up to $\sim$ 250 % – they nevertheless remain substantially smaller than the fractional variation of the continuum. For all three lines, the amplitude of the variability is comparable to but slightly larger than in NGC 5548 during the 1988-1989 campaign ($F_{var}$ = 0.17, 0.13 and 0.11, respectively; Clavel et al. 1991). On the other hand, the fractional variation of the C IV line in NGC 4151 over 6 years, 0.39 (Clavel et al. 1987), is significantly larger than in NGC 4593.

Rapid variability of Ly$\alpha\lambda$1216 took place in February 1985, from JD 6119 to JD 6123. The line flux decreased by a factor 1.8±0.15 in 96 hours ($\sim$ 4 days). This fade took place both in the line wing (highly contaminated by N V$\lambda$1240; Paper 1) and in the core (where N V is negligible) by factors 1.9±0.5 and 1.7±0.2 respectively, therefore suggesting that both Ly$\alpha$ and N V experienced rapid fluctuations. Note that during the same interval, the 1447 Å continuum also faded, by a factor 1.6. However, the decline of the continuum had already started on JD 6115, 4 days before that of Ly$\alpha\lambda$1216. This rapid variation yields a two-folding time $\Delta t_{\times 2}$ = 5±1 days for Ly$\alpha\lambda$1216.

## 4.2 Line-Continuum correlations

Figure 8 shows the intensity of the three strongest lines plotted as a function of the 1447 Å continuum flux. Table 4 lists the correlation coefficients, together with the corresponding confidence level.
Ly$\alpha\lambda$1216 and C IV$\lambda$1549 clearly correlate with the continuum. The trend may not however be linear, the line flux saturating at large values of the continuum intensity. Note that delays in the response of the lines due to the finite propagation time of the ionizing photons will tend to blur line–continuum correlations. The correlation is slightly better for Ly$\alpha\lambda$1216 than for C IV$\lambda$1549. Remarkably, the C III]$\lambda$1909 flux does not correlate with $F_{1447}$.

Similarly in the case of the FES, the overall strong line-continuum correlation, together with the very short time scale of variability of $F_{1447}$ precludes the existence of a large ($\geq$ 6 days) delay in the response of the lines to the continuum variations. The February 1985 light-curves show a well defined local maximum both in the UV continuum and in Ly$\alpha$ flux, allowing this idea to be tested further. Inspection of Fig 7 conveys the impression that Ly$\alpha\lambda$1216 lags behind the continuum by a few days. The maximum of the cross-correlation ($r = 0.85$) is shifted by +2 days toward positive lags (fig. 9). Again, the behaviour of both the core (maximum, $r = 0.8$, at +3 days) and the red wing ($r = 0.9$ at +2 days) are quite similar to the total line flux. The formal uncertainty on the delay (Gaskell & Peterson 1987) is ±1 day, which almost certainly underestimates the true error. To be conservative, we adopt $\Delta t \leq 4$ days for the delay of the Ly$\alpha\lambda$1216 line.
The cross-correlation of the C IV$\lambda$1549 line in February 1985 does not show a significant peak. Nevertheless, the behaviour of C IV$\lambda$1549 on other occasions is consistent with the hypothesis that its variations are closely related to those of the continuum but significantly more delayed than those of Ly$\alpha\lambda$1216: for instance, during the February 2–5, 1984 interval, while both Ly$\alpha\lambda$1216 and $F_{1447}$ decreased, C IV$\lambda$1549 remained constant. On the other hand, C IV$\lambda$1549 as well as Ly$\alpha\lambda$1216 followed the continuum decline of May 7–16, 1981. Taken at face value, this suggests $3 \leq \Delta t \leq 9$ days for the delay of C IV$\lambda$1549.

## 4.3 Line Ratios

In this section, we study the average values of the three main emission line ratios: Ly$\alpha\lambda$1216/C IV$\lambda$1549, Ly$\alpha\lambda$1216/Mg II$\lambda$2798 and C IV$\lambda$1549/C III]$\lambda$1909 as well as their temporal behaviour.

All three ratios show highly significant variations (Table 3, bottom). The average value of the Ly$\alpha$/C IV ratio is 1.4$\pm$0.2, where the error quoted represents the r.m.s. deviation of individual measurements about the mean. Such a ratio is significantly smaller than that found in an average Seyfert 1 galaxy ($\sim$ 2, Wu, Boggess & Gull 1983) or quasar (2.8$\pm$1.2, Wilkes 1986). The Ly$\alpha$/Mg II ratio in NGC 4593, 4.5$\pm$1.0, is also smaller than in other AGN's ($\simeq$ 8, Wu et al. 1983). On the other hand, the mean C IV/C III] ratio, 4.8$\pm$1.0, is fairly normal for a Seyfert 1 galaxy.

Only Ly$\alpha$/Mg II shows a significant correlation with the continuum (Table 4). However, because the Mg II$\lambda$2798 flux stays constant, this correlation is simply a consequence of that found earlier between Ly$\alpha\lambda$1216 and $F_{1447}$.

For the optical lines, we use the ratio of the means because of the small number of measurements. The Balmer decrements: H$\alpha$/H$\beta$=3.4$\pm$0.4 and H$\beta$/H$\gamma$=2.8$\pm$0.5, are both larger than the case B recombination values, (2.8 and 2.1 respectively at T=$10^4$ K, Osterbrock 1974), but smaller than the average ratio for a typical Quasar or Seyfert 1 (H$\alpha$/H$\beta$ ranges from 4-6, Netzer 1990). The Ly$\alpha$/H$\beta$ ratio, 6.8$\pm$2.0, is also smaller in NGC 4593 than in other Seyfert 1's or Quasars (8–15, Netzer 1990), but agrees well with the average value, 6.83, obtained by Wu et al. (1983) for a sample of 9 Seyferts (including NGC 4593). The ratio BaC/H$\beta\lambda$4861 = 26.6$\pm$4.0 is much larger than in other Seyfert 1s, *e.g.* 4.7 to 9.5 for the sample of WNW and 12 as a mean value for NGC 5548 (Wamsteker et al. 1990). Its is also larger than the theoretical range (1.5–15) spanned by the models of Kwan& Krolik (1981).

The average ratio of the total (UV+optical) Fe II flux to Ly$\alpha\lambda$1216 is Fe II/Ly$\alpha$ = 2.4$\pm$0.8, similar to that found by WNW in their sample of strong Fe II emitting Seyfert 1 galaxies and quasars. As noted by these authors, such a large value is difficult to reconcile with the standard photoionization model of the BLR as it requires much more flux to heat the transition region than is available. The Fe II(total)/Mg II$\lambda$2798 ratio, 10.9$\pm$2.3, is not significantly different from that in WNW's sample (8). More peculiar is the small value of the UV to optical Fe II line flux ratio, 1.66$\pm$0.42, (*cf* 4 to 12 in the WNW sample).

To summarize, the emission line spectrum of NGC 4593 is characterized by (i) a relative deficiency of Ly$\alpha\lambda$1216 photons, (ii) fairly intense Balmer continuous emission and (iii) low ionization lines (H$\beta\lambda$4861, Mg II$\lambda$2798 and Fe II) which are strong relative to Ly$\alpha\lambda$1216. All those facts are indicative of a very optically thick emitting medium. Ly$\alpha$/C IV is unusually small while C IV/C III] is fairly "normal".

### 4.4 Line profiles

The UV line profiles have been analysed in the appendix of paper 1.

Table 3 shows that the C IV$\lambda$1549 line profile is clearly asymmetric, since its blue wing is on the average 50% stronger than its red wing.

Both the wings and the core of Ly$\alpha\lambda$1216 and C IV$\lambda$1549 experienced significant variations (Table 3). The amplitude is larger in the wings than in the core. The evidence becomes stronger if one sums the flux in the blue and red wings of C IV$\lambda$1549. The fractional variation of the C IV$\lambda$1549 wings, 0.34, is much larger than that of the core, 0.20. Moreover, the wings exhibit rapid variability, with a two-folding time $\Delta t_{\times 2}$ = 8±2 days, whereas the core only varies on much longer time scales ($\Delta t_{\times 2}$ = 100±20 days). Furthermore, the C IV$\lambda$1549 wings correlate well with $F_{1447}$, whereas the core does not. On the other hand, the red wing and the core of Ly$\alpha\lambda$1216 correlate equally well with the UV continuum.

It is worth pointing-out that only Ly$\alpha\lambda$1216, C IV$\lambda$1549 and the He II lines display very broad wings extending up to 7000-10000 km s$^{-1}$ FWHM (paper 1). The wings of the lower ionization lines such as C III]$\lambda$1909, Mg II$\lambda$2798 and in particular the strong well exposed Balmer lines do not extend beyond 4400 km s$^{-1}$FWHM.

To summarize, the higher ionization lines have broader wings, vary with a greater amplitude and more rapidly than the lower excitation features. Moreover, the C IV$\lambda$1549 line wings vary more and more rapidly and correlate better with the continuum than the line core.

## 5. IMPLICATIONS

### 5.1 The continuum source in NGC 4593

#### *5.1.1. Where is the "big-bump" ?*

NGC 4593 is unusual in that its energy distribution extends with roughly the same spectral index $\alpha = -0.85$ from 1.2 $\mu$m up to 1 $keV$. The extrapolation of this power law to the hard X-rays agrees within a factor of 2.3±0.2 with the observed flux, in contrast to other AGN's where it over-predicts the $keV$ flux by 1 to 2 orders of magnitude. In other words, NGC 4593 apparently lacks the so-called "big-bump" which characterizes the SED of Seyfert 1's and quasars (Malkan & Sargent 1982; Malkan 1983).

The only hint of a spectral curvature over the 1 $keV$–1 $\mu$m range is the existence of a steep excess of soft radiation above the hard X-ray power-law. One possibility is that the soft excess represents the high energy tail of a "big-bump" which is otherwise too "hot" to radiate in the Ultraviolet. Since the work of Shields (1978), it has become customary to explain the "big-bump" as thermal emission from an accretion disk around a massive black-hole. If the disk is geometrically thin and optically thick (Shakura & Sunyaev 1973) a complete analytical solution can be obtained for the temperature profile. In particular, at 4.7 Schwarzschild radii, the disk reaches a maximum temperature:

$$T_{max} = 2 \times 10^5 \epsilon^{-0.25} (\frac{L}{L_{Edd}})^{0.25} (\frac{M}{10^6 M_\odot})^{-0.25} \ (\mathrm{K})$$

where M is the mass of the black-hole, $\frac{L}{L_{Edd}}$ the bolometric luminosity expressed as a function of the Eddington limit, and $\epsilon = 16^{-1}$, the efficiency of conversion of matter into energy around a Schwarzschild Black-Hole. Under the assumption that the emission line gas is gravitationally bound and using 3 lt-d – the average of the Ly$\alpha$ and C IV lag – as a measure of its radial distance, one can estimate the mass of the black hole in NGC 4593. It is first necessary to assign a characteristic velocity to the gas. Assuming that macro-turbulent motion of the BLR clouds is the main line broadening mechanism (Penston et al. 1990), the de-projected most probable turbulent velocity of the gas relates in a simple way to the line width: $V_{turb}$ = 0.6 FWHM. The full-width at half-maximum of the Ly$\alpha\lambda$1216 emission line is 3200 km $s^{-1}$ (while it mostly reflects the core velocity dispersion, it is safe to use it here as both the core and the wing vary with similar amplitude and delay - section 4 -). Therefore we infer M = $2.2^{+1.4}_{-1.1} \times 10^6$ $M_\odot$. Integrating the average X-ray flux from 0.1 to 100 $keV$ and the mean UV-optical power-law spectrum from 1 eV to 100 eV, one obtains $L = 3.3 \times 10^{43}$ erg.$s^{-1}$. This implies that NGC 4593 radiates at 12 % of its Eddington limit and, thus, the above equation yields $T_{max}$ = 197,000 K.

Electron scattering in the disk atmosphere raises this temperature by a factor $\simeq 2.5$ (Czerny & Elvis 1987) so that, in this approximation, the bump peaks near 210 eV (59 Å), *i.e.* within the range of values allowed by the spectral fits to the soft X-ray excess (paper 1). Assuming that the bulk of the radiation is released at 5 Schwarzschild radii and that the disk is seen face-on, its total luminosity is $L_{disk}$ = $2.8 \times 10^{42}$ erg.$s^{-1}$. The corresponding monochromatic flux received at earth, 8.2 $\mu$Jy at 0.24 $keV$, is comparable to that of the soft excess, 10$\mu$Jy on the average. Assuming $F_\nu \propto \nu^{1/3}$ spectral dependance, it emits only 0.003 mJy at 1447 Å, *i.e.* less than 0.3 % of the observed flux. Therefore, a hot thin accretion disk could account both for the soft excess and for the absence of "big bump" in the ultraviolet. This peculiar situation arises because the central mass in NGC 4593 is small, as illustrated by the fact that the emission lines are at the same time relatively narrow and rapidly variable. A small black-hole mass is also consistent with the low luminosity of NGC 4593.

The strongest line ratios are also consistent with the idea that the "Big Bump" is shifted toward high frequencies in NGC 4593. Given the ionization potentials of hydrogen (13.6 eV) and those of the $C^{2+}$ and $C^{3+}$ ions (47.9 and 64.5 eV, respectively), the Ly$\alpha\lambda$1216 line is driven solely by the extension of the UV power-law into the Lyman limit, while the C III]$\lambda$1909 and C IV$\lambda$1549 line fluxes presumably also depend in part on the intensity of the soft X-ray "big-bump". Even higher energy quanta are needed to penetrate and heat the transition region where the low excitation/ionization lines are thought to be formed. Hence, in the "standard" BLR model, the Mg II$\lambda$2798 and Fe II lines are effectively controlled by the 600-800 eV continuum band (Krolik & Kallman 1988). A soft X-ray "big-bump" associated with a steep UV power-law produces a relative deficit of photons near 13.6 eV. Hence, it naturally accounts for the fact that the Ly$\alpha$/C IV, Ly$\alpha$/Mg II and Ly$\alpha$/Fe II ratios are unusually low while the C IV/C III] ratio is "normal" in NGC 4593.

### *5.1.2. What is the origin of the Optical to X-ray power-law ?*

The fact that the ultraviolet and the optical flux vary in phase and in strict proportion one to the other suggests that a common underlying emission mechanism drives their flux fluctuations. Besides, the rough correlation between the UV and the hard X rays on long time scales suggest some kind of causal connection between the two bands. On the other hand, however, the observed two-folding times imply that the "radius" of the rapidly variable hard X-ray source is at most 1 lt-hours and that of the ultraviolet source less than 3 lt-ds. This argues against the possibility that the X-ray and UV sources are spatially coincident.

The situation is reminiscent of that found in NGC 5548 and NGC 4151. In both sources, significant correlations have been found between the hard X-rays and ultraviolet bands on time scales of a few days to several years (Clavel et al. 1992; Perola et al. 1986). However, in neither of the two galaxies do the rapid ($\sim$ hours) X-ray fluctuations have counterparts in the UV light-curve, where the shortest time for a factor two variation is $\geq$ 3 days. Clavel et al. have interpreted this result in terms of a thermal reprocessing model. The motivation for this model initially came for completely independent reasons, namely the need to explain the flattening of the X-ray spectrum above $\sim$ 10 $keV$ together with the large equivalent widths of the fluorescent Fe K$\alpha$ line near 6.5 $keV$, a common property of Seyfert 1 galaxies (Nandra, Pounds & Stewart 1991). Pounds et al. (1990) and later George & Fabian (1991) proposed a model where the X-rays source lies above and irradiates a cold disk or slab. Part of the flux ($\leq$ 10%) is Compton scattered back to the observer, while $\sim$ 90 % is absorbed. If the disk is cold ($\sim$ *a few* $10^5$ K), as required by the relatively low energy of the Iron line, the heat deposited must necessarily come-out as thermal radiation in the ultraviolet band. Hence, the "reflection" model implicitly predicts the existence of correlated variations of the X-ray and UV flux. The correlation does not need to be perfect, however. Indeed, a perfect correlation would imply an implausibly fine tuning, with the light travel time of the X-ray photons matching closely the variability timescale of the X-ray source. A very attractive feature of the reprocessing model is that it naturally explains the quasi-simultaneity of the variations of the UV and optical flux, in NGC 4593 as well as in NGC 5548

and NGC 4151.

Could the reprocessing scenario apply to NGC 4593 as well ? As stated above, the observational situation is qualitatively the same in NGC 5548 and NGC 4593 (and NGC 4151) – although the correlation between the X-rays and the UV seems to be weaker for the second AGN, as explained above in this scenario one does not expect a strong correlation to hold between them. In particular, the best-fit line between the 2–10 $keV$ and UV flux goes through the origin in both objects. As noted by Clavel et al. (1992), this implies that the UV variations are completely driven by the X-rays and therefore that the whole ultraviolet emission originates from thermal reprocessing of the X-ray flux. Is that energetically possible ? The available mean X-ray luminosity integrated from 1 to 100 $keV$ is $1.6 \times 10^{43} erg\ s^{-1}$, sufficient to account for the average 1 to 100 $eV$ UV-optical flux $1.3 \times 10^{43} erg\ s^{-1}$. The two-folding time of the 1447 Å continuum corresponds to a dimension of 7.6 $10^{15}$ cm, or $11,600\ R_s$ for a black-hole mass of $2 \times 10^6 M_\odot$. Hence, the UV emitting region corresponds to the periphery of the accretion disk. In a thin disk, the heat generated internally by viscous dissipation of the angular momentum is completely negligible in these outermost regions. Can external heating by X-rays reconcile the disk model with the NGC 4593 observations ? If the height at which the X-ray source is located is commensurate with the dimension of the disk (e.g. the radial distance across the disk, R), the amount of energy deposited per unit surface area falls-off as $R^{-2}$. Hence, the effective temperature of the disk drops as $R^{-0.5}$, rather than the usual $R^{-0.75}$ without reprocessing. In the black-body approximation, this results in a $F_\nu \propto \nu^{-1}$ power-law spectrum, very close to what is observed. In this scenario, the soft X-ray excess originates from the innermost layers of the disk, where the gas temperature reaches $\sim$ 200,000 K, as discussed above. It draws its energy mainly from the heat generated internally by viscous dissipation of the gravitational energy. Local thermal instabilities (with time scales ranging from 0.8 to 8 hours for a $2{\times}10^6 M_\odot$ black hole) will trigger variations of the soft X-ray flux which are independent from those of the hard X-ray source. Hence, in this model, one would not expect the hard and soft-excess X-ray flux to vary in phase. The absence of a correlation between the two wavebands is consistent with such a scheme and may in fact imply that external irradiation is negligible in the energy balance of the innermost layers of the disk. This would be consistent with the above hypothesis that the hard X-ray source lies far above the disk surface.

To summarize, the thermal reprocessing models explains several distinctive features of NGC 4593: (1) the steepness of its UV-optical spectrum, (2) the existence of a weak correlation between its hard X-ray and ultraviolet flux variations, (3) the correlated variations of its ultraviolet and optical flux, (4) the quasi-simultaneity of the variations in the two bands and (5) the absence of correlated hard and soft X-ray variations.

The peculiarity of NGC 4593, as compared to other AGN's, arises from its comparatively small black-hole mass, $2 \times 10^6 M_\odot$, versus 3 and $4 \times 10^7 M_\odot$ in NGC 4151 and NGC 5548, respectively (Clavel et al. 1991 and Clavel et al. 1992). On the other hand, the dimensions of its ultraviolet and X-ray sources are comparable to those in the other two objects. Since its X-ray luminosity is also comparable to that of NGC 4151, external irradiation by hard X-rays reaches much further in NGC 4593, out to a region where viscous heating is negligible and the energy balance of the disk is completely dominated by X-rays. This results in an unusually steep UV spectrum, the usual "big blue bump" being shifted toward EUV and soft X-ray frequencies.

Are there viable alternative to the reprocessing model ? One possibility is that the optical-UV spectrum has a non thermal origin. Synchrotron self-Compton (SSC) models such as those of Band & Grindlay (1986) and Zdziarski (1986) can successfully match the IR to X-ray spectrum of AGN's. They predict a connection between the radio and UV/optical flux as well as significant degree of polarization in the UV, optical and near IR continuum. Connecting the NGC 4593 6 cm flux (1.6 mJy) with its 1447 Å data-point requires a power-law of spectral index $\alpha = 0.0$. It seems therefore unlikely that the UV-optical $\alpha = -0.85$ continuum is the optically thin synchrotron tail of the radio spectrum. Martin et al. (1983) measured the polarization in white light (3800-5600 Å) through a $4''$ aperture of a sample of 99 Seyfert galaxies. For NGC 4593, they obtain p=0.52±0.16 %, close to the amount of interstellar polarization expected from the hydrogen column density on the line-of-sight, (p=0.27 % Serkowski et al. 1975). A tiny amount of polarization by interstellar grains inside the host galaxy of NGC 4593 could easily reconcile the two values. Hence, both the observed low frequency energy distribution and the absence of intrinsic polarization in the optical probably rule out the "Blazar" model as a viable explanation for the origin of the SED in NGC 4593.

### *5.1.3. Hot dust in the near-IR ?*

The pattern of variability of the NGC 4593 near IR flux is reminiscent of that of the bright Seyfert 1 galaxy F 9 (Clavel et al. 1989). In this AGN, the maximum of the IR variability occurs in the K band, and the amplitude decreases with increasing wavelength. Clavel et al. (1989) argued that the origin of the variable near IR flux in F 9 was thermal reprocessing of the UV flux by hot dust grains close to their sublimation temperature. The finding that the variations of the H, K and L flux in F 9 are delayed with respect to those in the far ultraviolet and that the delay corresponds closely to the light travel time of the UV photons to the dust evaporation radius lent considerable credence to their model.

To check the applicability of the dust scenario to NGC 4593 one can compare the shortest time scale of variability with that expected from grains at $\sim 1500$ K, the dust sublimation temperature ($T_{sub}$). The rapid variations of February 1985 suggests that the dimension of the near IR source in NGC 4593 is at most $37 \pm 2$ lt-d at K. This is small, but consistent with the dust emission model. The mean ultraviolet luminosity of NGC 4593 in February 1985, integrated from 10 to 100 eV is $5.4 \times 10^{42}$ erg.s$^{-1}$. This corresponds to a dust evaporation radius of $36 \times (T_{sub}/1500K)^{-2.8}$ lt-d (Barvainis 1987), in good agreement with the observational upper limit on the size of the near IR source. In a dust model, one expects the longest wavelength IR radiation to vary less and more slowly than the shorter wavelength flux, simply because the latter originates in a region which is hotter, more compact and closer to the UV source than the region responsible for the lower frequency radiation. Such a pattern is indeed observed in NGC 4593 (see section 3.1.4. above).

Following the calculations of Clavel et al. (1989), it is possible to estimate the mass of dust responsible for the variable near IR emission. At a distance of $37 \pm 12$ lt-ds, the grain equilibrium temperature is $T_{gr} = 1440 \pm 150$ K, where the quoted error takes into account a 50 % uncertainty on the mean integrated 10–100 eV luminosity. For the average grain discussed in Barvainis (1987) and Clavel et al. (1989), one needs $(5 \pm 2) \times 10^{-4} M_{\odot}$ of dust to account for the average net 2.2 $\mu$m flux. If the dust is spread in a spherical shell whose thickness is of the order of its radius, then the shell will produce a visual extinction $\sim 0.25$ magnitudes. It is more likely that the hot dust region represents the inner edge of the putative molecular torus which is thought to surround the nucleus of AGN's. As pointed out by Clavel et al. (1989) for Fairall 9, there must be a much bigger reservoir of cool dust at larger radial distances in order to account for the observed mid to far IR flux.

Rapid near IR variability is observed in two other Seyfert 1's NGC 1566 (Baribaud et al. 1992) and NGC 3783 (Glass 1992) where it is also consistent with the dust reprocessing model. It is interesting that in these two objects as well as in Fairall 9 and NGC 4593, the dust shell is more commensurate with the scale of the outermost part of the broad line region than with that of the NLR, *i.e.* it appears to lie at the expected location of the inner edge of the molecular torus.

## 5.2 The broad line region in NGC 4593

Most emission lines are variable and their variations correlate with those of the UV continuum. This lends strong support to the hypothesis that the BLR gas is heated and photoionized by the continuum source.

The absence of a correlation between lines like C III]$\lambda$1909 or the core of C IV$\lambda$1549 and the continuum, can easily be explained by light travel time effects. The fact that Ly$\alpha\lambda$1216 and C IV$\lambda$1549 (particularly its wings) correlate with $F_{1447}$, on the other hand, is consistent with the idea that they originate from a compact region close to the continuum source. Given the observed two-folding time of $F_{1447}$, these lines presumably originate from a region whose size is of the order of 6 lt-d or less.

A similar dimension was found for the C IV wing emitting region in NGC 4151 by Clavel et al. (1990).

The rapid, large amplitude variations of Ly$\alpha\lambda$1216 give an independent argument in favour of an upper limit of $\sim$ 5 lt-days to the size of the emitting region. Its quick response ($2 \pm 2$ days) to the continuum variations is consistent with this limit and implies that the bulk of the Ly$\alpha\lambda$1216 line emitting material is confined within 4 lt-d from the continuum source.

The amplitude of the variations increases along the sequence Mg II- C III]-C IV-Ly$\alpha$(+N V). This suggests that the BLR is stratified, *i.e.* that the bulk of the highly ionized gas is confined within a region more compact than the assembly of clouds which have an extensive neutral or partially ionized zone. A similar pattern is observed in NGC 5548 (Clavel et al. 1991).

The fact that the C IV$\lambda$1549 wings vary with larger amplitudes and correlate better with the continuum than the line core further suggests the existence of a velocity gradient in the BLR, *i.e.* that the bulk of the high velocity material lies closer to the ionizing source than the lower velocity gas. Support for these ideas also comes from the fact that only the high ionization lines such as Ly$\alpha$, C IV$\lambda$1549 and the He II features show extended wings. A similar but better established gradient was found in NGC 4151 (Clavel et al. 1990). A velocity gradient is consistent with a model where the gas is gravitationally bound to a central black-hole.

Both the relative small value of Ly$\alpha$/H$\beta$ and the unusually large BaC/H$\beta$, are indicators of very optically thick BLR clouds. The small value of the Fe II(UV)/Fe II(opt) ratio is a further indication of large optical depths in the transition zone of the BLR clouds.

The BLR covering factor, $\epsilon$, can be estimated by comparing the number of Ly$\alpha\lambda$1216 photons and the number of ionizing photons (integrated over the 1 ryd to 300 eV range). One obtains $\epsilon \approx 0.14$.

The emerging BLR picture is that of a geometrically thick distribution of clouds where the gas is heated and photoionized by the continuum and where the physical conditions varies with the distance to the ionizing source. This prompted us to perform photoionization model calculations in order to try to reproduce the characteristics of the spectrum and derive the run of physical conditions as a function of radial distance. Details of the models and discussion of the results are given in the appendix. To a first order approximation, a two-zone model accounts successfully for the UV spectrum of NGC 4593. It is characterized by an inner zone of radius 3 lt-d, fairly dense ($10^{11}$ cm$^{-3}$) and thin (column density $N_H = 10^{21}$ cm$^{-2}$), and another zone at 15 lt-d, less dense ($10^{10}$ cm$^{-3}$) and thicker ($N_H = 10^{23}$ cm$^{-2}$). However this model fails to account for the large strength of the Fe II and Balmer lines.

Is it possible to boost the intensity of the Balmer lines relative to Ly$\alpha$ and C IV? Collin-Souffrin (1987), Collin-Souffrin & Dumont (1990) and Dumont & Collin-Souffrin (1990a,b) have developed a model where the bulk of the low ionization lines flux originate from the periphery of an accretion disk. At large radial distances, viscous dissipation inside the disk is insufficient to maintain an adequate temperature for efficient line emission. Hard X-rays directly illuminating the disk from above and depositing their energy in its upper layers provide the extra heat needed. Because the optical depth is very large (infinite) and the density quite high ($\geq 10^{12}$ cm$^{-3}$), the resonance lines are completely thermalized, the inter-combination lines are collisionally suppressed and the cooling flux can only leak through lines originating from metastable levels. The emerging spectrum is dominated by Fe II, Ca II and Balmer lines and continuum emission. Such a model could provide the missing Balmer line flux and account for the intense BaC and Fe II line observed in NGC 4593. This picture is also consistent with the continuum reprocessing scenario outlined above. Another possibility is that dust grains *internal* to the BLR clouds absorb part of the UV emission. A small amount of dust is very efficient at destroying trapped Ly$\alpha\lambda$1216 photons because of the large optical depth in that line. This is not likely however, since at the distance inferred for the clouds, dust grains would evaporate. To survive, the dust would have to be located outside the Strömgren zone of each cloud, in its extended partially ionized back-side. An additional difficulty with this scenario is that it cannot explain the observed large Balmer continuum to H$\beta\lambda$4861 intensity ratio.

To conclude, theoretical calculations reinforce our earlier conclusion that the BLR in NGC 4593 is stratified, since it is impossible to model its spectrum with a single homogeneous slab: A model with at least two zones at 3 and 15 lt-d is required to explain the observations. It is likely that Fe II, Balmer line and continuum are emitted in a distinct medium which we identify with the periphery of the irradiated accretion disk. More observational data are needed to further test this idea and better constrain the model.

## 6. Summary

The multi-frequency monitoring campaign of the Seyfert 1 galaxy NGC 4593 yielded the following results:

1. The near IR to far UV energy distribution is well described by a single, unusually steep ($\alpha = -0.85 \pm 0.05$) power-law. Hence, NGC 4593 lacks a so-called "big-bump" which dominates the optical and ultraviolet spectrum of nearly all other Seyfert 1's and quasars. Above $\sim 2\ keV$ the spectrum flattens, its index becoming $\alpha = -0.75 \pm 0.05$.

2. The only deviation from the IR to X-ray power-law occurs in the soft X-ray range where there is a marked up-turn of the spectrum. This soft X-ray excess has a steep ($\alpha \leq -2.6$) spectrum and varies with an amplitude which is larger than that of the IR to X-ray power-law or the hard X-ray component.
3. The amplitudes of the variations - defined as the ratio of the maximum to the minimum measured flux - are $3.19 \pm 0.08$, $5.75 \pm 0.33$, $4.0 \pm 1.1$, $3.0 \pm 0.6$ and $3.2 \pm 0.6$ at 4 $keV$, $\sim 0.5$ $keV$, 1447 Å, $\sim$ 5000 Å, and 1.65 $\mu$m, respectively.
4. The continuum is rapidly variable in all wavebands, from the near IR to the hard X-rays implying that the emitting source is unusually compact. The two-folding time scales are 1.1 hour in the 2–10 $keV$ band, 70 hours at 1450 Å and $37 \pm 12$ days at 2.2 $\mu$m.
5. To the level of accuracy of our measurements, the IR to X-ray power-law retains a constant spectral shape while its flux varies. However, we cannot rule-out the existence of subtle variations of the spectral index of order $\Delta\alpha = -0.27$ or less.
6. The variations of the UV and optical continuum correlate with each other. Given that the UV flux occasionally doubles in less than 3 days, this further implies that the optical continuum cannot lag behind the far UV by more than 6 days. The hard X rays and the UV fluctuations are also correlated, though the correlation has only marginal significance. Similar but better established correlations have been obtained for NGC 5548 (Clavel et al. 1992) and NGC 4151 (Clavel et al. 1991).
7. Neither the weak X-ray/UV correlation nor the absence of a detectable lag between the optical and ultraviolet can be accommodated in the framework of a geometrically thin accretion disk model. Rather, they suggest that the bulk of the UV and optical flux arise from thermal reprocessing of the hard X-rays which originate from above the disk and irradiate its upper layers.
8. The high excitation Ly$\alpha\lambda$1216, C IV$\lambda$1549, C III]$\lambda$1909 and He I$\lambda$5876 emission lines are strongly variable, with amplitudes of $2.56 \pm 0.26$, $2.41 \pm 0.28$, $2.61 \pm 0.74$ and $3.1 \pm 0.4$ respectively. On the other hand, lines from less ionized species or arising from lower excitation levels – such as Mg II$\lambda$2798, Fe II or H$\beta\lambda$4861– remain constant, at least to the level of accuracy of our measurement (10 and 20 % respectively). The Balmer continuum flux – prominent in NGC 4593 – is also constant.
9. The variations of the high excitation emission lines correlate with those of the UV continuum and the linear correlation coefficient decreases along the sequence Ly$\alpha\lambda$1216, C IV$\lambda$1549, Si IV$\lambda$1394, C III]$\lambda$1909. This lends strong support to the idea that at least the high excitation lines arise from photoionization of the BLR gas by the UV continuum source.
10. The Ly$\alpha\lambda$1216 line lags behind the UV continuum by at most 4 days. This, together with the fact that the two-folding time of the line is $5.1 \pm 0.9$ days, indicates that the bulk of the Ly$\alpha\lambda$1216 flux arises from a very compact region, of the order of a few light-days. The dimension of the line emitting region in NGC 4593 is comparable to that found in NGC 4151 (Clavel et al. 1990), an AGN of similar luminosity.

11. The fact that the high ionization/excitation lines are strongly variable whereas the low excitation lines are not, together with the fact that the degree of correlation with the continuum decreases with the level of excitation of the line suggests a stratified picture of the BLR where the highly ionized/excited gas is confined within a smaller volume than the gas which is neutral or partially ionized.

12. The high excitation lines are very broad, with wings extending beyond $\sim$ $10,000\,kms^{-1}$ in the case of C IV$\lambda$1549, in contrast to the lower excitation features which are significantly narrower ($\sim$ $3000\,kms^{-1}$ FWHM). Moreover, the wings of the C IV$\lambda$1549 line ($|V| \geq 3000\,kms^{-1}$) vary with a larger amplitude and more rapidly than its core ($|V| \leq 1500\,kms^{-1}$). This strongly suggests the existence of a velocity gradient in the BLR where the velocity dispersion of the clouds increases toward small radii.

13. Under the assumption that the emission line gas is gravitationally bound, the delay of Ly$\alpha\lambda$1216 together with its width can be used to derive the mass of the black-hole in NGC 4593, $2 \times 10^6 M_\odot$. Such a small mass is consistent with the low bolometric luminosity of this galaxy and implies that NGC 4593 radiates at 12 % of its Eddington limit.

14. Because the black-hole mass is relatively small, the accretion disk is significantly hotter in NGC 4593 than in other AGN's, so that its "big-bump" is shifted into the EUV. For a $2 \times 10^6 M_\odot$ black-hole, the maximum temperature of the disk, 200,000 K, and its luminosity are consistent with the parameters inferred from a spectral fit to the soft X-ray excess.

15. The two-folding time of the 2.2 $\mu$m continuum, 37$\pm$12 days, corresponds closely to the light-travel time of the UV photons to the dust evaporation radius. This together with the fact that the near IR variability is maximum in the H and K band strongly suggests that the few micron flux originates from thermal radiation by hot ($\sim$ 1400 K) dust grains.

## APPENDIX A: Photoionization model calculations

We have used the photoionization code CLOUDY (version 76–03, June 1990; Ferland 1990, Ferland & Rees 1988) in order to infer the BLR physical parameters in NGC 4593 when comparing with the observed spectrum. The necessary inputs to the model which can be inferred directly or indirectly from the present observations are the spectral shape and the luminosity of the continuum and, in some cases, the average distance of the gas to the ionizing source. Some words of caution are necessary, however. First, we only have a rough qualitative idea of the shape of the continuum in the all too important EUV range. Second, the only clue to the radial distance of the gas is the delay in the response of the Ly$\alpha\lambda$1216 emission line to the variations of the continuum. This delay yields an estimate of the BLR emissivity weighted "radius" valid for Ly$\alpha\lambda$1216 only. However, since there are strong indications that the BLR in NGC 4593 is stratified, one has to consider a range of distances. In view of the discussion in section 5.2, reasonable guesses for this parameter are bracketed by 3 and $\sim$ 20 lt-d.

The incident continuum used in these computations represents the average of the observed flux in each wavelength band, after de-reddening and correction for stellar light contamination (see paper 1). For the radio band, we have used the data of Ulvestad & Wilson (1984). The far IR fluxes are those measured by IRAS. The near IR data-point are the average of our J, H, K and L measurements. The optical-UV is the $\alpha = -0.85$ power law, normalized to the mean $F_{1447}$ and the X-rays are given by a $\alpha = -0.75$ power law, normalized to the mean amplitude (as obtained in the fits to the EXOSAT data, *i.e.* $n_0$ in table 2 of paper 1).The UV power law is extrapolated up to 300 eV, as suggested by the soft X ray excess spectral fits (table 4 in paper 1). Following Edelson & Malkan (1986) or Engargiola et al. (1988), the radio component is extended up to 0.5 cm (5000 $\mu$m) and the far IR is assumed to break down at 100 $\mu$m. The interpolation of the fluxes at this two frequencies gives a 1082 $\mu$m flux of 41 mJy and a spectral index for the low energy tail of the IR $\alpha = +2.07$, in agreement with the observations of Lawrence et al. (1991) ($F_{1082\mu m} = 31 \pm 22$ mJy and $\alpha > +1.76$).

The free parameters of the models are the column density of an individual cloud ($N_H$), the hydrogen density, n, (or gas pressure) and the cloud distance to the ionizing source. The cloud covering factor comes as an output when forcing the total line intensities to match the observed ones. The ionization parameter is also univoquely determined by the observed total luminosity and the assumed density and distance.

A grid of (r, $N_H$, n) models was computed, r ranging from 3 to 20 lt-d, $N_H$ from $10^{21}$ to $10^{25}$ cm$^{-2}$ and n from $10^9$ to $10^{12}$ cm$^{-3}$.

The very weak or unobserved $C^+$, $N^+$ or $O^0$ lines set stringent upper limits on the column density. At density lower than $n = 10^{11} cm^{-3}$, the upper limits on the C II]$\lambda$2326, N II]$\lambda$2140 and O I$\lambda$1304 intensities relative to Ly$\alpha\lambda$1216 or Mg II$\lambda$2798 rule out all models with $N_H \geq 10^{23} cm^{-2}$. The observed upper limit on the C II$\lambda$1335 to Ly$\alpha\lambda$1216 intensity ratio further rules out models with densities of $10^{12}$ $cm^{-3}$ or higher.

The intensities of the UV lines which lack very broad wings and those of the Ly$\alpha\lambda$1216, C IV$\lambda$1549 and He II line cores are well reproduced with clouds located at 15 lt-d from the continuum source, quite dense ($10^{11}$ $cm^{-3}$) and thin (column density $N_H$=$10^{21}$ $cm^{-2}$). On the other hand, this model fails to reproduce the correct UV to optical lines ratios: while the theoretical Balmer decrement matches the observations (H$\alpha$/H$\beta$=3.4 and H$\beta$/H$\gamma$=2.8), the theoretical Ly$\alpha$/H$\beta$ and Mg II/H$\beta$ line ratios are larger than the mean observed values by a factor of 2.3$\pm$0.7 and 5$\pm$1, respectively. The model also leaves unexplained the strong Fe II line flux as well as the intense Balmer continuous emission which characterize NGC 4593.

The very broad Ly$\alpha$, C IV and He II line wings can be explained if they arise in clouds closer to the central source (3 lt-d), less dense (n=$10^{10}$$cm^{-3}$), and with larger column densities ($N_H$=$10^{23}$$cm^{-2}$)). The sum of the spectra emerging from the 2 regions is shown in table A1, together with the average observed UV spectrum. The inferred covering factors are 0.06 for the 3 lt-d region and 0.17 for the 15 lt-d zone.

Table 1: Continuum variability parameters

| Continuum | Mean$^{(a)}$ | $F_{var}$ | R | $\chi^2_\nu$/dof | $\Delta t^{(b)}_{\times 2}$ |
|---|---|---|---|---|---|
| X rays, 4KeV | $1.77\pm0.78\,10^{-3}$ | 0.432 | 3.19±0.08 | 788.4/6 | 1.1 h |
| X rays, Al P | 0.051±0.034 | 0.664 | 4.78±0.54 | 219.6/5 | ··· |
| X rays, lexan | 0.13±0.07 | 0.565 | 5.75±0.33 | 627.7/7 | ··· |
| Soft excess lexan | 0.041±0.040 | 0.892 | ≥5.4±1.3 | 16.79/6 | ··· |
| $F_{1333\mathring{A}}$ | 0.84±0.37 | 0.439 | 4.13±0.82 | 36.74/19 | 3.1±0.9d |
| $F_{1447\mathring{A}}$ | 0.94±0.43 | 0.436 | 4.0±1.1 | 22.55/19 | 2.3±0.6d |
| $F_{1726\mathring{A}}$ | 1.36±0.59 | 0.414 | 4.0±1.1 | 20.67/19 | (3.3±1.3d) |
| $F_{1803\mathring{A}}$ | 1.45±0.57 | 0.387 | 3.88±0.60 | 50.01/19 | ··· |
| $F_{2011\mathring{A}}$ | 1.46±0.82 | 0.237 | 3.5±1.5 | 1.57/16 | ··· |
| $F_{2227\mathring{A}}$ | 2.27±0.54 | 0. | 2.59±1.54 | 0.687/16 | ··· |
| $F_{2710\mathring{A}}$ | 3.53±0.71 | 0.190 | 1.92±0.18 | 10.99/16 | ··· |
| $F_{3031\mathring{A}}$ | 4.00±0.71 | 0.141 | 1.94±0.21 | 3.20/16 | ··· |
| FES(obs) | 22.3±2.0 | 0.082 | 1.36±0.06 | 7.603/20 | ··· |
| FES(nuc) | 5.80±2.0 | 0.316 | 2.96±0.58 | 7.607/20 | ··· |
| $J_{obs}$ | 57.8±2.3 | 0.030 | 1.159±0.044 | 2.38/21 | ··· |
| $H_{obs}$ | 81.7±5.1 | 0.056 | 1.318±0.050 | 5.59/21 | ··· |
| $K_{obs}$ | 83.2±7.5 | 0.086 | 1.355±0.053 | 11.86/21 | ··· |
| $L_{obs}$ | 89.±12. | 0.125 | 1.66±0.11 | 8.77/19 | ··· |
| $J_{nuc}$ | 13.1±2.3 | 0.133 | 2.01±0.40 | 2.38/21 | ··· |
| $H_{nuc}$ | 21.7±5.1 | 0.212 | 3.16±0.63 | 5.59/21 | ··· |
| $K_{nuc}$ | 36.9±7.5 | 0.194 | 2.16±0.21 | 11.86/21 | (37±12d) |
| $L_{nuc}$ | 63.±12. | 0.177 | 2.07±0.20 | 8.77/19 | (34±17d) |

(a) All means in mJy, except the soft X-rays in counts, and reddening corrected ($E_{B-V}$=0.029)

(b) Two-folding times; in parenthesis if the flux changed by a factor less than 2 (but ≥ 1.5)

Table 2: Continuum–continuum correlations

| Features | | | Corr.Coeff. | P($r \geq$) | $\chi^2_\nu$ |
|---|---|---|---|---|---|
| ME | vs | $F_{1447}$ | 0.924 | 0.025 | 0.82 |
| LE | vs | $F_{1447}$ | 0.970 | 0.006 | 0.67 |
| Soft excess | vs | $F_{1447}$ | 0.600 | 0.290 | ··· |
| $F_{1803}$ | vs | $F_{1447}$ | 0.980 | $8\ 10^{-12}$ | 0.71 |
| $F_{1803}$ | vs | $F_{1333}$ | 0.930 | $8\ 10^{-10}$ | ··· |
| $F_{2710}$ | vs | $F_{1447}$ | 0.896 | $4\ 10^{-8}$ | 1.6 |
| FES | vs | $F_{1803}$ | 0.853 | $1.8\ 10^{-6}$ | 1.9 |
| FES | vs | $F_{1447}$ | 0.806 | $1.8\ 10^{-5}$ | 1.8 |
| J | vs | H | 0.878 | $7.8\ 10^{-8}$ | 0.39 |
| H | vs | K | 0.946 | $3.0\ 10^{-11}$ | ··· |
| K | vs | L | 0.830 | $6.0\ 10^{-6}$ | ··· |
| J | vs | L | 0.757 | $1.1\ 10^{-4}$ | ··· |

Table 3: Line variability parameters

| Feature | Mean$^{(a)}$ | $F_{var}$ | R | $\chi^2_\nu$ $^{(b)}$ | $\Delta t_{\times 2}$ $^{(c)}$ |
|---|---|---|---|---|---|
| Ly$\alpha$1216 | 258±72 | 0.275 | 2.56±0.26 | 45.7 | 5.1±0.9 |
| CIV$\lambda$1549 | 180±42 | 0.226 | 2.41±0.28 | 27.8 | (17±4) |
| CIII]$\lambda$1909 | 39±9 | 0.195 | 2.61±0.74 | 2.91 | (6±3) |
| MgII$\lambda$2798 | 57±5 | 0 | 1.50±0.34 | 0.382 | ... |
| MgII fit | 55±8 | 0.098 | 1.71±0.32 | 1.78 | ... |
| SiIV$\lambda$1400 | 24±10 | 0.342 | 5.1±3.5 | 6.49 | 4±3 |
| NIV]$\lambda$1485 | 13±7 | 0.125 | 13 ± 40 | 1.08 | ... |
| HeII+OIII] | 19±9 | 0.168 | ≤5 ± 7 | 1.35 | ... |
| FeII UV | 362±88 | 0.105 | 2.12±0.57 | 1.39 | ... |
| BaC | 1030±130 | 0 | 1.58±0.45 | 0.417 | ... |
| FeII m42 | 62±7 | 0 | 1.3±0.6 | 0.22 | ... |
| HeII$\lambda$4686 | 11±2.8 | 0.178 | 1.8±0.5 | 1.48 | ... |
| H$\beta$ | 38±3 | 0 | ... | 0.71 | ... |
| HeI$\lambda$5876 | 18±10 | 0.576 | 3.1±0.4 | 52 | ... |
| H$\alpha$ | 130±10 | 0 | ... | ... | ... |
| Ly$\alpha_{core}$ | 47.±12. | 0.254 | 2.48±0.29 | 23.82 | (6±1) |
| Ly$\alpha_{red}$ | 52.±17. | 0.302 | 3.01±0.74 | 13.01 | (4.5±1.9) |
| CIV$_{core}$ | 45.7±9.6 | 0.200 | 2.31±0.27 | 16.83 | ... |
| CIV$_{red}$ | 23.7±8.9 | 0.268 | 4.1±2.2 | 2.56 | (7±6) |
| CIV$_{blue}$ | 37.±16. | 0.393 | 3.9±1.6 | 6.39 | 6±2 |
| CIV$_{core_tot}$ | 85.±17. | 0.195 | 2.21±0.21 | 24.07 | ... |
| CIV$_{blue+red}$ | 60.±22. | 0.341 | 3.3±0.8 | 7.40 | 8±2 |
| Ly$\alpha$/CIV | 1.42±0.18 | 0.107 | 1.86±0.24 | 4.62 | ... |
| Ly$\alpha$/MgII | 4.54±1.05 | 0.228 | 2.52±0.20 | 41.36 | ... |
| CIV/CIII] | 4.76±0.99 | 0.205 | 2.03±0.12 | 39.67 | ... |
| Ly$\alpha$/CIV$_{core}$ | 1.03±0.13 | 0.076 | 1.71±0.24 | 6.975 | ... |
| Ly$\alpha$/CIV$_{red}$ | 2.23±0.51 | 0 | 2.20±0.70 | 1.947 | ... |

(a) All fluxes in units of $10^{-14}$erg s$^{-1}$cm$^{-2}$ and corrected for reddening ($E_{B-V}$=0.029)

(b) Significant variability (at the 99% level) requires minimum $\chi^2_\nu$ of 1.9 for the UV lines except the Fe II and MgII (2.1) and 3.3 for the optical lines, except HeI (4.5)

(c) Two folding times in days

Table 4: Line–continuum correlations

| Features | | | Corr.Coeff. | Conf.level % |
|---|---|---|---|---|
| Ly$\alpha\lambda$1216 | vs | $F_{1447}$ | 0.791 | 99.997 |
| C IV$\lambda$1549 | vs | $F_{1447}$ | 0.709 | 99.95 |
| C III]$\lambda$1909 | vs | $F_{1447}$ | 0.365 | (89) |
| Si IV$\lambda$1394 | vs | $F_{1447}$ | 0.686 | 99.91 |
| Ly$\alpha\lambda$1216/C IV$\lambda$1549 | vs | $F_{1447}$ | 0.453 | (95.5) |
| Ly$\alpha\lambda$1216/Mg II$\lambda$2798 | vs | $F_{1447}$ | 0.735 | 99.88 |
| C IV$\lambda$1549/C III]$\lambda$1909 | vs | $F_{1447}$ | 0.431 | (94.) |
| Ly$\alpha\lambda1216_{core}$ | vs | $F_{1447}$ | 0.768 | 99.992 |
| Ly$\alpha\lambda1216_{red}$ | vs | $F_{1447}$ | 0.758 | 99.989 |
| C IV$\lambda1549_{core}$ | vs | $F_{1447}$ | 0.534 | (98) |
| C IV$\lambda1549_{red}$ | vs | $F_{1447}$ | 0.582 | (99.2) |
| C IV$\lambda1549_{blue}$ | vs | $F_{1447}$ | 0.653 | 99.8 |
| C IV$\lambda1549_{blue+red}$ | vs | $F_{1447}$ | 0.703 | 99.95 |
| Ly$\alpha\lambda1216_{red}$ | vs | Ly$\alpha\lambda1216_{core}$ | 0.905 | 100 |
| C IV$\lambda1549_{red}$ | vs | C IV$\lambda1549_{cor}$ | 0.731 | 99.97 |
| Ly$\alpha\lambda$1216/C IV$\lambda1549_{core}$ | vs | $F_{1447}$ | 0.682 | 99.91 |
| Ly$\alpha\lambda$1216/C IV$\lambda1549_{red}$ | vs | $F_{1447}$ | 0.212 | ($\ll$85.) |

Table A1: Comparison between observed spectra and photoionization models

| Line[a] | Observed | Model |
|---|---|---|
| | $10^{-14}$erg s$^{-1}$cm$^{-2}$ | |
| Ly$\alpha\lambda$1216+N V$\lambda$1240 | 260±70 | 210 |
| O I$\lambda$1304 | 4±1 | 0.2 |
| C II$\lambda$1335 | 2±1 | 10 |
| Si IV$\lambda$1394+O IV$\lambda$1400 | 24±9 | 19 |
| N IV]$\lambda$1486 | 13±7 | 2 |
| C IV$\lambda$1549 | 180±40 | 140 |
| He II$\lambda$1640+O III]$\lambda$1663 | 19±9 | 32 |
| N III$\lambda$1750 | 7±2 | 1.5 |
| Al III$\lambda$1860 | 2±1 | 4.7 |
| C III]$\lambda$1909+Si III]$\lambda$1892 | 39±9 | 41 |
| N II]$\lambda$2140 | $\leq$ 0.9 | 0.2 |
| C II]$\lambda$2326 | $\leq$ 0.5 | 1 |
| Mg II$\lambda$2798 | 57±5 | 42 |

(a) All lines $\geq 10^{-3}\times$Ly$\alpha$ shown

## Figure Captions

Figure 1.: The ultraviolet continuum flux at 1803 Å as a function of the 1447 Å flux, both corrected for reddening.

Figure 2.: The nuclear spectral energy distribution of NGC 4593. The X-rays are represented by the mean best-fit power-law. The UV optical and near IR data-points are from the simultaneous observations of February 15, 1985. The best-fit UV-optical power-law is also shown. The arrow in the upper-left corner indicates the level of the radio flux.

Figure 3.: The de-reddened continuum light-curves at 1447, 2710 and 5200 Å (FES).

Figure 4.: The left panels show the complete K (top) and L (bottom) light-curves. The February 1985 light curves are expanded in the right panels.

Figure 5.: The X-ray versus UV (1447 Å) flux. Top: the 2–10 $keV$ ME band and bottom: the soft X-rays LE band.

Figure 6.: The optical FES count rate (5200 Å flux) as a function of the UV continuum flux at 1447 Å.

Figure 7.: Excerpts from the light curves of the 1447 Å continuum (a), the Ly$\alpha\lambda$1216 emission line (b) and C IV$\lambda$1549 (c) during episode of intense monitoring. Note that the time axis is not continuous. Only those epochs when 2 or more observations had been obtained within a few days are shown.

Figure 8.: The Ly$\alpha\lambda$1216 (top), C IV$\lambda$1549 (middle) and C III]$\lambda$1909 (bottom) emission line intensities as a function of the UV continuum flux at 1447 Å.

Figure 9.: The cross correlation of the Ly$\alpha\lambda$1216 emission line with the 1447 Å continuum in February 1985 (heavy line) together with the auto-correlation of the continuum (thin line).